\documentclass[10pt]{dis03}
\usepackage{epsf}

\newcommand{\tvec}[1]{{\mathbf{#1}}}

\begin{document}

\title{\hfill \parbox{9em}{{\mdseries
	DESY-03-125 \\ hep-ph/0309070 \\ \\ \\}} \\
DIFFRACTION, EXCLUSIVE PROCESSES, \\ 
AND PROTON STRUCTURE IN THREE DIMENSIONS\thanks{{}Talk given at the
${\rm 11^{th}}$ International Workshop on Deep Inelastic Scattering
(DIS 03), St.~Petersburg, Russia, 23--27 April 2003.  To appear in the
proceedings.}}

\author{M.~DIEHL \\
Deutsches Elektronen-Synchroton DESY,
22603 Hamburg, Germany}

\maketitle

\begin{abstract}
\noindent I briefly review how a three-dimensional picture of parton
dynamics is rendered in the impact parameter representation, applied
to generalized parton distributions and to the dipole picture of
high-energy scattering.
\end{abstract}


\section{The impact parameter representation}

Processes like deeply virtual Compton scattering, elastic meson
production, and hard diffraction tell us about the spatial
distribution of quarks or gluons within a hadron.  The impact
parameter representation provides an adequate language to represent
this information.  It is a deep concept in field theory and has been
used in various contexts of high-energy scattering.

To describe fast moving particles (hadrons or partons) it is useful to
work with light-cone momenta $p^\pm = p^0 \pm p^z$.  {}From the usual
momentum eigenstates of a particle---labeled here by $|p^+,
\tvec{p}\rangle$ with boldface vectors referring to the $x$-$y$
plane---one can form states
\begin{equation}
\label{impact-state}
|p^+, \tvec{b}\rangle = (16\pi^3)^{-1} 
	\int d^2 \tvec{p}\, e^{-i \tvec{p} \tvec{b}}\,
	|p^+, \tvec{p}\rangle
\end{equation}
with definite plus-momentum and definite transverse position (called
impact parameter).  To avoid infinities one can use wave packets,
where the transverse position is smeared out a bit.  To retain the
simple interpretation of $p^+$ as roughly twice $p^z$ (for $p^z>0$)
one must restrict the range of $\tvec{p}$ in the wave packet, giving a
position uncertainty $\delta b \gg 1 / p^+$ \cite{Diehl:2002he}.  One
can hence localize a particle in the $x$-$y$ plane very precisely,
provided it moves fast enough along $z$.  This is different from
trying to localize a particle in \emph{all three} space dimensions,
where one is limited by the Compton wavelength $1/m$.  The original
work of Burkardt \cite{Burkardt:2000za} gives a hint at how this
difference arises: it turns out that to achieve a simple
interpretation one must approximate the energy $p^0$ with a constant
for a given wave packet.  To localize in three dimensions this
restricts momenta to $(p^x)^2 + (p^y)^2 + (p^z)^2 \ll m^2$.  For
localizing in the transverse plane one needs $(p^x)^2 + (p^y)^2 \ll
m^2 + (p^z)^2$, which is not restrictive for large $p^z$.

The states $|p^+, \tvec{b}\rangle$ are eigenstates of a transverse
position operator $\tvec{R}$ \cite{Soper:1972xc}.  It is associated
with ``transverse boosts'', i.e., with Lorentz transformations
\begin{equation}
\label{transv-boost}
p^+ \to p^+ , 
\qquad\qquad
\tvec{p} \to \tvec{p} - p^+ \tvec{v}
\end{equation} 
for some fixed vector $\tvec{v}$.  Notice the similarity with Galilean
transformations in non-relativistic mechanics, where the mass of a
particle replaces $p^+$.  Noether's theorem relates Galilean
invariance with the conservation of the center of mass of a many-body
system.  The conserved quantity for transverse boost invariance is
hence the ``center of (plus) momentum'', $\sum p^+_i
\tvec{b}_i^{\phantom{+}} / \sum p^+_i$.  The position $\tvec{b}$ of a
hadron state (\ref{impact-state}) can be understood as the center of
momentum of its partons.


\section{Parton distributions in impact parameter space}

The impact parameter representation fits naturally with the concept of
parton distributions, as realized long ago by Soper
\cite{Soper:1977jc} and shown in detail by Burkardt
\cite{Burkardt:2000za,Burkardt:2002hr}.  This can be used for
``imaging'' of hadrons \cite{Ralston:2001xs}.  The usual momentum
space parton densities can be expressed through matrix elements
$\langle p^+, \tvec{p} |\, \mathcal{O} \,| p^+, \tvec{p}\rangle$,
where the quark or gluon operator $\mathcal{O}$ creates and
annihilates a parton at impact parameter $\tvec{0}$.  I have omitted
the dependence on polarization labels and on the usual plus momentum
fraction $x$ of the parton in the target.  Inserting the same operator
between the states (\ref{impact-state}) one finds
\begin{equation}
\label{parton-imp}
\langle p^+, \tvec{b}' \,|\, 
	\mathcal{O} \,| p^+, \tvec{b}\rangle 
\propto \delta^{(2)}(\tvec{b} - \tvec{b}')  
\int d^2 \tvec{\Delta}\, e^{i \tvec{b} \tvec{\Delta}}\,
\langle p^+, \tvec{p}' \,|\, 
	\mathcal{O} \,| p^+, \tvec{p}\rangle ,
\end{equation}
where the matrix element on the right-hand side depends on $\tvec{p}$
and $\tvec{p}'$ only through $\tvec{\Delta} = \tvec{p}' - \tvec{p}$
because of Lorentz invariance.  The expression on the right is the
Fourier transform of a generalized parton distribution
\cite{Muller:1994fv} at $t = - \tvec{\Delta}^2$ and zero skewness
$\xi$, whereas the left-hand side gives the joint density of partons
with plus momentum fraction $x$ and transverse position $\tvec{b}$.
It hence contains genuinely \emph{three-dimensional} information about
the target structure.  Integrating (\ref{parton-imp}) over $\tvec{b}$
one recovers the usual parton densities with their one-dimensional
information.  If one integrates over $x$ then $\mathcal{O}$ turns into
a local operator, and the right-hand side becomes the Fourier
transform of a form factor associated with this operator.  This
representation of a form factor in terms of a two-dimensional density
is fully relativistic, unlike the usual representation through a
three-dimensional density in a hadron at rest, which holds in the
nonrelativistic limit and is limited to distance scales large compared
with the Compton wavelength of the target.  This is not a big
restriction for a large and heavy nucleus, but it is quite restrictive
for a proton.  An interpretation of generalized parton distributions
along these lines has been proposed in \cite{Ji:2003ak}.

\begin{figure}
\begin{center}
  \leavevmode 
  \epsfxsize=0.65\textwidth 
  \epsfbox{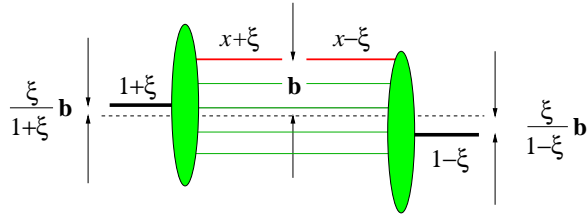}
\caption{\label{fig:impact} Impact parameter representation of a GPD
in the region $x\ge \xi$.  Plus momentum fractions $1\pm \xi$ and $x
\pm \xi$ refer to the average of initial and final proton momenta.
The dashed line indicates the origin of transverse coordinates.}
\end{center}
\end{figure}

It is not known how to measure generalized parton distributions at
$\xi=0$ (except for certain moments in $x$), but the processes
mentioned in the beginning involve these functions at nonzero $\xi$.
A calculation analogous to (\ref{parton-imp}) gives the picture shown
in Fig.~\ref{fig:impact} for $x\ge \xi$, and a similar one for $x\le
\xi$ \cite{Diehl:2002he}.  The center of momentum of the hadron is
shifted by the interaction because the struck parton returns with a
smaller momentum fraction.  Note that this shift is independent of
$x$, which is integrated over in the amplitudes of hard scattering
processes.  In a wide range of kinematics the shift is small compared
with the distance of the struck parton from the ``average'' hadron
position.

\section{The dipole representation}

\begin{figure}[hb]
\begin{center}
  \leavevmode 
  \epsfxsize=0.73\textwidth 
  \epsfbox{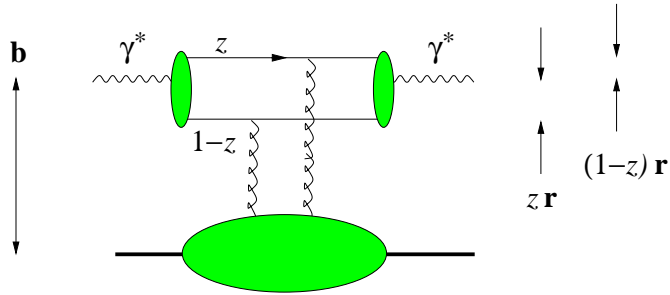}
\caption{\label{fig:dipole} Impact parameter representation of the
virtual Compton amplitude $\gamma^* p \to \gamma^* p$ in the dipole
picture.}
\end{center}
\end{figure}

A well-known formulation of BFKL dynamics at leading $\log
\frac{1}{x}$ accuracy is the color dipole representation, shown in
Fig.~\ref{fig:dipole} for the virtual Compton amplitude.  The sum of
momentum space Feynman graphs can be rewritten in terms of photon wave
functions $\Psi(z,\tvec{r})$ and of the scattering amplitude
$N(x_{bj}, \tvec{r}, \tvec{\Delta})$ for a $q\bar{q}$ dipole of size
$\tvec{r}$ on the target.  The size $\tvec{r}$, which is Fourier
conjugate to the transverse quark momentum in the loop, is conserved
in the scattering process as a nontrivial consequence of small-$x$
kinematics, see e.g.\ \cite{Bialas:2000xs}.  After Fourier transform
with respect to the transverse momentum transfer $\tvec{\Delta}$ one
obtains a representation fully in impact parameter space as shown in
the figure.  The conservation of the center of momentum in the
transition between photon and $q\bar{q}$ pair has direct consequences
for the dipole scattering amplitude, which is of the form
\cite{Bartels:2003yj}
\begin{equation}
\widetilde{N}(x_{bj}, \tvec{r}, \tvec{b}) =
  \int d^2 \tvec{\Delta}\, 
  e^{i \tvec{\Delta}\, [\tvec{b} + (1-z) \tvec{r}]}
  N(x_{bj}, \tvec{r}, \tvec{\Delta}) .
\end{equation}
We see that the relevant transverse distance in the Fourier transform
is between the target and the quark in the dipole (not between the
target and the photon).  If one finally takes the leading double $\log
\frac{1}{x} \log Q^2$ approximation, the amplitude $\widetilde{N}$
becomes proportional to the generalized gluon distribution at impact
parameter $\tvec{b}$, with the factorization scale $\mu \sim 1 /|
\tvec{r} |$ set by the size of the dipole probing the gluons.

\section{Conclusions}

The impact parameter formalism gives a geometrical representation of
hadron structure and dynamics, fully consistent with the principles of
relativistic field theory.  It naturally blends with the physical
picture of the QCD parton model.  Various aspects of physics become
quite transparent in this representation, such as Gribov diffusion and
diffractive shrinkage \cite{Gribov:1973jg,Bartels:2000ze}, the pion
cloud of the nucleon \cite{Strikman:2003gz}, or the limit $x\to 1$ of
a fast parton in the target \cite{Burkardt:2002hr}.  Much remains to
be understood, as illustrated by the nontrivial interplay of
$\tvec{r}$ and $\tvec{b}$ in saturation dynamics \cite{Stasto:2003di}.
Measurement of the $x_{bj}$ and $t$ dependence in exclusive $ep$
scattering processes can provide correlated information in longitudinal
momentum and transverse position space, and thus in all three space
dimensions.


\end{document}